# Fundamental Limitations in Advanced LC Schemes

A.A. Mikhailichenko

*Cornell LEPP, Ithaca, NY 14850*

**Abstract**. Fundamental limitations in acceleration gradient, emittance, alignment and polarization in acceleration schemes are considered in application for novel schemes of acceleration, including laser-plasma and structure-based schemes. Problems for each method are underlined whenever it is possible. Main attention is paid to the scheme with a tilted laser bunch.

## OVERVIEW

The presence of particles with energy >1000 TeV in cosmic showers tells us that these high-energy particles were somehow involved in a process of formation of our Universe. At the high edge of the spectrum, single particle can carry up to 1 kJ of energy. Greisen-Zatsepin-Kuz'min predicted that cosmic rays with energies over $5 \times 10^{10}$ GeV would interact with cosmic microwave background photons producing pions via the $\Delta$ resonance $p + \gamma \rightarrow \Delta \rightarrow p + \pi$. This process continues until the cosmic ray energy falls below the pion production threshold (*GZK* limit) [1]. On the other hand, equivalent Lab frame energy for a future 1TeVx1TeV linear collider corresponds to fixed target experiment carried out at $E_{Lab} = E_{cm}^2 / mc^2 \cong 8 \cdot 10^9$ GeV, which is only ten times lower than GZK limit. This is a challenging inspiration for Advanced Concept Accelerators.

The only way to answer the question about the involvement of particles with high energy in processes running in the Universe is to create them in the Lab by acceleration with technologies available for this purpose. Schemes with a laser as a source of radiation are interesting for us. These schemes do not suggest any media between the laser radiation and the particle (so we exclude all plasma methods from the candidates). But some *agent* between the laser radiation and the particle is required by fundamental physics laws.

## ACCELERATION AS A TWO-PHOTON PROCESS

In lowest order, the electron-photon interaction can be described by a two-photon Feynman diagram. The presence of a second (radiated) photon allows, for example, particle acceleration by a plane wave; the process of acceleration is going on while the particle radiates. In terms of photon absorption, the cross section of this process decreases with energy, preventing usage of this method at high energy. Particles acquire many RF photons during the process of acceleration in a cavity. If one could make a source of photons having *TeV* energy, then electron acceleration could be arranged by a single act of photon absorbtion with the radiation. The same requirement for the presence of the second photon can be expressed through consideration of energy balance in a classical description of the process of energy acquisition while a particle passes through the cavity. The acceleration term appears as a *product* of the wake field and the field excited in the cavity by an external RF source. Of course, all photons excited in the cavity are coherent. Change of angular momentum is defined by the formula

$$\Delta L = \frac{1}{c^2} \int \vec{r} \times ((\vec{E}_0 + \vec{E}_{sp}) \times (\vec{H}_0 + \vec{H}_{sp})) dV - \frac{1}{c^2} \int \vec{r} \times (\vec{E}_0 \times \vec{H}_0) dV = mc\gamma\vartheta, \qquad (1)$$

where $E_0, H_0$ stand for the electric and magnetic fields excited by the RF generator, $E_{sp}$, $H_{sp}$ stand for the spontantenous fields excited by the particle, so the cavity can deflect (focus) particles running in an appropriate phase.

Summarizing, the acceleration structure serves for delivery of the second photon in Feynman's diagram – spontaneous radiation (wakes). The accelerating structure also serves for *confinement* of EM field in space. Its precise location is defined by the accuracy of fabrication, accuracy of positioning and by how far from equilibrium the fields are. So the structure can not be much larger than the wavelength of the laser radiation; otherwise the fluctuations in the process of establishing the fields will generate long living (in terms of period) perturbations with undesirable spatial structure. That is why so-called photonic structures are useless for particle acceleration with short RF pulses. One good property of a small structure is that it cannot accommodate thermal photons, so there is no interaction between the beam and thermal photons–an analog of the GKZ effect with thermal photons [2].

It is always interesting to find out *where the second photon is hidden in plasma acceleration schemes*, but this topic will be not touched here.

Proper orientation of the accelerated field is also important. Namely, let us consider the PASER [3] as an example. PASER is a system that uses an active (in the sense of Laser activity) medium. The active medium ($CO_2$ gas at ~0.25 atm) is excited by discharge of an electrical pulse, so the atoms of this medium are changed to an excited state. When the electron beam passes through, its electric field triggers a transition of excited atoms into the ground state, while radiated photon are absorbed by the electrons. However, the electric field of the relativistic bunch is *transversely polarized*. So the induced radiation will "pump" the transverse field, which increases the transverse momenta, not the longitudinal one. At any moment, the resulting force is due to the induced radiation defined by the averaging of momenta transferred to the field by many atoms. So this system is not able to accelerate a relativistic beam. In original experiments of Leypunksy and Latyshev [4], the electrons were nonrelativistic, so the active medium could increase the temperature of the electron beam. In some sense, this is acceleration, but not in the sense required for high energy physics application. Although the energy of a particle can grow, it is associated with the growth of the transverse momenta only. That is why a PASER concept cannot work for acceleration to high energy.

In contrast with structure-based schemes, in the plasma-driven schemes the accurate EM field positioning is not possible. This is caused by fluctuations of the charge forming the field boundary, because the charge density in a plasma, ~$10^{16}$/cm$^3$, is much smaller than the electron charge density in conducting materials, ~$10^{23}$/cm$^3$. In a dilute plasma with a density of ~$10^{-6}$ the density in a metal, the ratio of Debye radius to the wavelength becomes $r_D/\lambda \cong 10^{-1}$ only; $r_D \cong \sqrt{k_B T/4\pi n e^2}$, $\lambda$ is the laser wavelength. The plasma medium cannot confine the EM field in a steady centered way, unlike the electron plasma in metals doing this for much larger wavelength. Therefore, the long-term acceleration in a plasma-driven acceleration schemes *is not possible*; (acceleration of positrons is also not realistic in plasma methods). In addition, it is hard to identify what is the beam emittance *at the exit* of a plasma-based accelerator.

So the fundamental requirements for any acceration system are:
- Cross interference of the accelerating and spontanenous (wake) fields.
- Spatial orientation of the accelerating field must be directed along the particle's trajectory.
- Centering of the accelerating field should not be sensitive to the fluctuations.

## TRAVELING LASER FOCUS (TLF)

One method fully satisfies all these requirements [5]. This method uses multi-cell microstructures. Each cell of the microstructure has an opening from one side. Focused laser radiation excites the cells by a special optical device through these openings locally, in accordance with instant position of accelerated micro-bunches of particles in the structure. A tilted (sloped) laser bunch can be prepared by two methods: by sweeping or with a grating.

### Realization of Traveling Laser Focus with a Sweeping Device

This method was suggested in 1989 [5]. It uses electrooptical prism crystals driven by a high-voltage pulse. The sweeping device can be characterized by the deflection angle $\vartheta$ and by the angle of natural diffraction – $\vartheta_d^s \cong \lambda/a$, where $a$ is the aperture of the sweeping device. The ratio of the deflection angle to the diffraction angle is a fundamental measure of the quality for any deflecting device. This ratio defines the *number of resolved spots*

*(pixels)* controllably positioned along the structure, $N_R = \vartheta/\vartheta_d$. The last number is an invariant under optical transformations. Electrooptical sweeping devices use the dependence of the refractive index on the electric field strength applied to some crystals. When an electric field is applied to such a crystal, the refractive index *n* changes its value. For a prism-based device, this yields a change in deflection angle. To arrange such a change, the sides of the prism are covered by metallic foils and a high voltage is applied to them. When *E(t)* is applied to the crystal, the refractive index changes $\Delta n = \Delta n(E(t))$, so the number of resolved spots $N_R$ becomes

$$N_R \cong \Delta\vartheta/(\lambda/a) \cong \Delta n \cdot l/\lambda, \qquad (2)$$

where *l* is the length of the prism device, $a \sim L$– is the aperture of the sweeping device. In order to increase these numbers ($N_R, \Delta\vartheta$), multiple-prism deflectors were developed. We recognized that for the angular sweep of a short laser bunch, the electric field should be applied as a traveling wave with a *slope*. Multi-prism deflectors have oppositely oriented optical axes in neighboring prismatic crystals. In this case, the full length of the deflecting device serves as *l*. For *l*=50cm, $w \cong 1$ cm, one can expect that the deflection angle can be $\Delta\vartheta \cong 10^{-2}$ and $N_R \cong 100$ for $\lambda \cong 1$ μm. Therefore, these numbers satisfy TLF method and there are no physical limitations here.

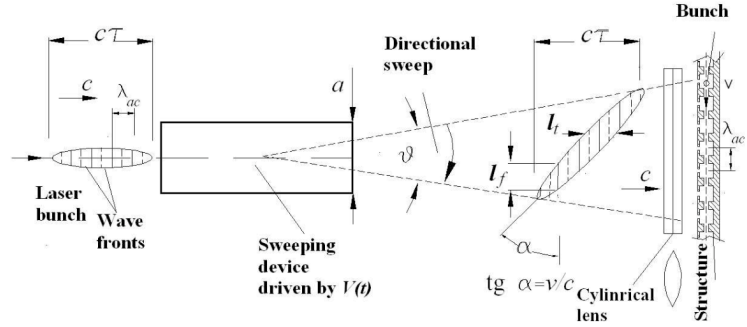

**FIGURE 1.** TLF principle of preparation of sloped laser bunch with the sweeping device [4]. Electron bunch is moving from the top to the bottom inside the structure.

In this case, the length of the laser bunch is about the length of the accelerating structure. The diffraction angle for the sweeping device is $\vartheta_d^s \cong \lambda_{ac}/a$. The lens installed in front of the sweeping device has its focal point on the structure. One can find a few examples of engineering for the sweeping device in Ref. [6].

## Realization of Traveling Laser Focus with Grating

The other method for creation of the sloped laser bunch makes use of a diffraction grating [7], [8]. First of all, the laser bunch must be short in this method with a duty time which is equal to the effective laser duty in the previous method, $l_t$. The pulse must have a width of about twice of the length of the accelerating structure, which can be arranged with appropriate telescopic optics. The incoming laser bunch hits the grating, which has a tilt of 63o with respect to direction of propagation. The effective length of the accelerating structure, chosen the same as in the previous section, is $L \cong c\tau$. The geometric relations are clear from Fig. 1. As it follows from the principle of operation of a grating, formation of reflection in necessary direction requires many periods and involves only a small area ~δ on the grating. This inevitably extends the laser pulse length. Really, the diffraction angle in this case is $\cong \lambda_{ac}/a$, where the area involved in formation of reflection chosen for comparison with the sweeping method as small as $\delta \approx l_f \cong l_t$, $\lambda_{ac}$ stands for the accelerating wavelength coinciding with the wavelength of laser radiation. For the sweeping device, we have $l_t \cong L/N_R \cong a/N_R$. So for comparison of these two schemes, we represent the diffraction angle as $\cong \lambda_{ac}/a$. The ratio of diffraction angles in these two methods goes as

$$\vartheta_d^g / \vartheta_d^s \cong \sqrt{\lambda_{ac} N_R/a} /(\lambda_{ac}/a) \cong \sqrt{N_R a/\lambda_{ac}}. \qquad (3)$$

With some optimization of the grating profile, this could be improved, probably, to $\vartheta_d^g/\vartheta_d^s \cong N_R$ at the best. Therefore, the advantage of using the sweeping device is obvious—it gives a much smaller laser spot size in the longitudinal direction. The difference is ~100 times the minimum in a favor of the sweeping device. Therefore, the sweeping method is the preferred one [11].

# ARRANGEMENT OF LONG-TERM ACCELERATION WITH TLF

Long-term acceleration with many structures can be arranged as shown in Fig. 2 [13]. Limitation of the electric field strength arising from the tunneling of electrons through the well is known in quantum mechanics. Taking into account the image potential, the diffusion coefficient becomes equal to unity for the electric field strength $\sim 10^8$ $eV/m$ [9]. So 10 $GeV/m$ is the limiting number for practical purposes. In any case, the final answer will be revealed in experiments with different materials.

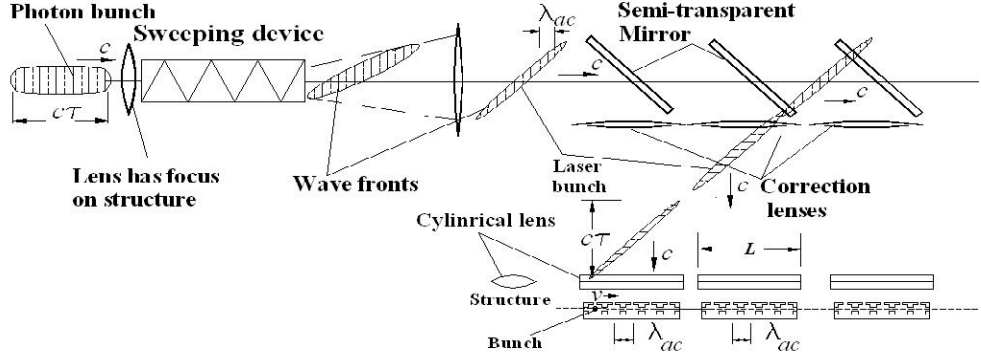

**FIGURE 2.** Long-term acceleration with the TLF method.

# MINIMAL EMITTANCE AND MAXIMAL POLARIZATION

With some exaggeration one can say that the laser acceleration business is a matter of the smallest emittance one can achieve for the injected beams. For an electron gas in a volume $V$, the total number of electrons in all states can be estimated for a uniform distribution as

$$N = \int dn \cong 2 \frac{p_x p_y \Delta p_\parallel}{(2\pi\hbar)^3} \frac{S_\perp l_b \gamma}{} \cong 2 \frac{\gamma\varepsilon_x \gamma\varepsilon_y \gamma l_b (\Delta p/p_0)}{(2\pi\lambdabar_C)^3} = 2 \frac{\gamma\varepsilon_x \gamma\varepsilon_y \gamma\varepsilon_z}{(2\pi\lambdabar_C)^3}, \qquad (4)$$

where $\gamma\varepsilon_z = \gamma l_b (\Delta p/p_0)$ –is an invariant longitudinal emittance, $l_b$ is the bunch length, $\gamma\varepsilon_x$ and $\gamma\varepsilon_y$ are the transverse horizontal and vertical emittances. Again, if $N$ is close to the number of the particles in the bunch, then the particles in the bunch become close to the degeneracy condition. We can say also that the beam with the number of the particles $N$ cannot have emittances lower than defined by this formula, namely $\gamma\varepsilon_x \gamma\varepsilon_y \gamma\varepsilon_z \geq \frac{1}{2}(2\pi\lambdabar_C)^3 N$, [10]. The physics is clear: all lower states are occupied. The minimal number of particles required to obtain the desired luminosity $L$ can be estimated as $N^2 \geq 4\pi\lambdabar_C^2 L/nf$, where $\lambdabar_C = \hbar/mc$, $f$ is a repetition rate, $n$ is the number of bunches per train. For $L \cong 10^{34} cm^{-2} s^{-1}$, $f$=100 $kHz$, $n$=10, $N \geq 4 \cdot 10^2$. We also mentioned in [12] that small emittance with reduced number of particles can be obtained by scraping all the extra particles obtained from the usual beam injectors.

One interesting circumstance here is that the *fully degenerate beam has zero polarization*, as each state is occupied by two electrons having oppositely oriented spins.

## Further Compression

A magnetic dipole $\vec{m}$ creates a magnetic field according to $\vec{H} = 3\vec{n} \cdot (\vec{m} \cdot \vec{n}) - \vec{m}/R^3$. In cases a) and b), Fig. 3, the unit vector $\vec{n}$ is orthogonal to the magnetic moment $\vec{m}$ associated with the electron/positron, so the magnetic field at the location of second electron is $\vec{H} = -\vec{m}/R^3$. In cases c) and d) the magnetic field is $\vec{H} = 2\vec{m}/R^3$, so this case is twice as effective for the field value. As the energy associated with magnetic moment of a pair of electrons is $E = \vec{m} \cdot \vec{H} = (3(\vec{m} \cdot \vec{n})^2 - \vec{m}^2)/R^3$, so the attracting force acting between two electrons due to theirs magnetic moments is

$$F \cong -\frac{\partial E}{\partial R} \cong 3\frac{3(\vec{m} \cdot \vec{n})^2 - \vec{m}^2}{R^4}, \qquad (5)$$

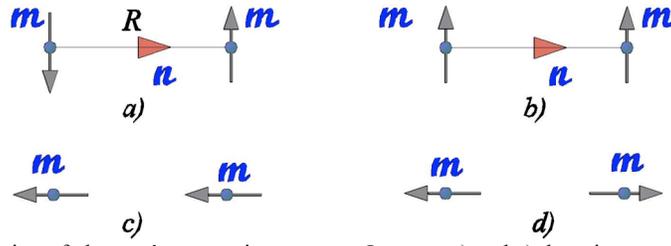

**FIGURE 3.** Relative orientation of electron's magnetic moments. In cases *a*) and *c*) there is attraction by the magnetic force; in cases *b*) and *d*)- repulsion.

which should be compared with the repulsive electric force $F \cong -e^2 / R^2$, so the balance is

$$3\frac{3(\vec{m} \cdot \vec{n})^2 - \vec{m}^2}{R^4} \cong \frac{e^2}{R^2} . \tag{6}$$

Substituting for the magnetic moment of electron $|\vec{m}| = e\hbar / 2mc$, one can show that the distance for which these two forces equalize each other is $R \cong \sqrt{3(2)} \hbar / 2mc = \sqrt{3(2)} \lambdabar_C / 2$, where the factor 2 in brackets corresponds to the case c). The energy required to bring two electrons to a distance equal to the Compton wavelength is $e^2 / \lambdabar_C \cong e^2 / r_0 \cdot \alpha \cong \alpha \cdot mc^2$, *i.e.* pretty small compared with the energy of transverse motion at the Interaction Point (IP).

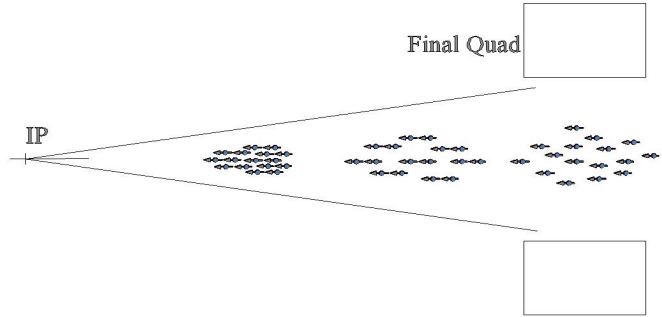

**FIGURE 4.** Clustering of a bunch of polarized electrons/positrons while it is moving towards IP.

So the polarized beams of positrons/electrons can serve to arrange the ultra-luminous condition. Minimal emittance which could be expected in this case comes to $\gamma\varepsilon_x \gamma\varepsilon_y \gamma\varepsilon_z \geq \frac{1}{2}(2\pi\lambdabar_C)^3$ i.e. as small as if electrons obey the Bose-Einstein statistics. Polarization of the bunch could be either one or zero. While colliding, the coupling between electron/positron pair destroyed by the incoming beam and collision with the opposite bunch is going as usual.

## SCALING IN A STRUCTURE-BASED ACCELERATOR

Transverse wakes scale *linearly* with spatial scaling, as follows from their dimensions: Volt/Coulombs/meter. Impedances are

$$Z_\parallel \cong -i\frac{Z_0}{R\sigma_b} \cdot (rh - ...) \to [Z_0], \quad Z_\perp \cong -i\frac{Z_0}{R^3} \cdot (rh - ...) \to Z_0[1/meter], \tag{7}$$

where $r$ is the radius of cavity counted from the drift tube, $h$ its height (along the beam trajectory), $R$ is the radius of the drift tube or iris, $Z_0 \cong 377$ Ohm. One can see that if the bunch length together with other dimensions just scales down as the wavelength, so that the amplitude of longitudinal current remains the same (as the charge density remains the same), then the voltage applied to the beam induced along the cavity will be the same, $V_\parallel \cong I_\parallel Z_\parallel = const$. The voltage applied to the beam across the cavity will also be the same, since despite the transverse impedance increased inversely proportional to the wavelength, the charge is reduced in the same proportion, $V_\perp \cong I_\perp Z_\perp = const$.

So basically, the structure is scaled down from 10 cm to 1 μm i.e. $10^5$ times. If we accept that the bunch population for a 10-cm cavity is ~$10^{10}$, then for the laser-scale cavity the bunch population should be ~$10^5$. The structure of our interest (the foxhole type [12], [13]) has a transverse gap, which makes the transverse wakes in the direction of the slit negligible. In the orthogonal direction, the slit focuses particles. So for the scaled structure, despite the fact that the number of cells is growing (for fixed longitudinal distance), the wavelength of betatron motion remains stable, as the misalignments of the neighboring cells can be treated as statistically independent. For focusing in the direction along the slit arranged by small quadrupoles see [13].

## SUMMARY

Limitations arising from the fundamental physical phenomena and processes should be taken into account when projecting accelerating installations with plasma and the structure-based ones. One class of limitations is associated with the nature of the electron/positron bunch as a Fermi-Gas. The sole occupancy by a single particle in each state immediately yields a limitation of minimal emittance and maximal polarization which could be achieved in the colliding beams. Magnetic force generated by magnetic moments allows clustering electrons/positrons in pairs with the distance between them of the order of the Compton wavelength. Now pairs, treated as bosons can condense further while moving towards IP.

Limitations in the density of carriers may restrict the precise alignment of the accelerating field due to fluctuations of the geometrical center of the accelerating fields in a plasma. Different capabilities of the plasma methods for focusing electrons and positrons during acceleration might be important as well.

For the structure-based schemes, we presented limitations in accelerating gradient arising from the quantum tunneling of electrons through a potential barrier, limiting achievable gradient to 1-10 *GeV/m* depending on the material of the structure. For acceleration scheme with a sloped laser beam, we considered limitations in preparation of such a beam by the laser sweeping device and by a grating. We observe that the sweeping device method is the preferred one.